# Gravitomagnetic London Moment and the Graviton Mass inside a Superconductor

C.J. de Matos[*]

*ESA-HQ, European Space Agency, 8-10 rue Mario Nikis, 75015 Paris, France*

M. Tajmar[†]

*ARC Seibersdorf research GmbH, A-2444 Seibersdorf, Austria*

**Abstract**

Using Proca electromagnetic and gravitoelectromagnetic equations the magnetic and gravitomagnetic properties of a rotating superconductor are respectively derived. Perfect diamagnetism, and the magnetic London moment are deduced from the photon mass in the superconductor. Similarly, it is shown that the conjecture proposed by the authors to resolve the cooper pair mass anomaly reported by Tate, can be explained by a graviton mass in the superconductor different with respect to its expected cosmological value.

[*] General Studies Officer, Phone: +33-1.53.69.74.98, Fax: +33-1.53.69.76.51, E-mail: clovis.de.matos@esa.int
[†] Head of Space Propulsion, Phone: +43-50550-3142, Fax: +43-50550-3366, E-mail: martin.tajmar@arcs.ac.at

**Introduction**

A superconductor exhibits two main magnetic properties. One is to expel any magnetic field from its interior (Meissner-Ochsenfeld effect) and the second one is to produce a magnetic field (London moment) when set into rotation. While the first property is well established and extensively verified, the latter one is still missing in many textbooks and is an area of active research. Already predicted by Becker et al[1] as a property of a perfect conductor, the London magnetic field is given by

$$\vec{B} = -2\frac{m^*}{e^*}\vec{\omega} , \qquad (1)$$

where $m^*$ and $e^*$ are the bare mass and charge of the Cooper-pairs, and $\omega$ the angular velocity at which the superconductor rotates. Several explanations can be found in the literature for the origin of this effect. The shortest heuristic derivation postulates that effective forces in the rotating system must vanish; the London moment is then needed to cancel the Coriolis force[2]. The earliest argument from Becker et al[1] considers the Cooper-pairs as a superfluid embedded frictionless in a positively charged ion matrix. If the superconductor is then rotated, the Cooper-pairs initially remain unaffected. This in turn causes an electric field, generated by induction due to the current of the positive ion matrix. Therefore, the Cooper-pairs will move with the ion matrix so that the electric field is zero. Only within the penetration depth, the Cooper-pairs will leg behind. This net current is then the source of the London moment. Maybe the simplest argument is that the canonical momentum of the Cooper-pairs must be zero (considering a superconductor that is much larger than the penetration depth)[3,4],

$$m^*\vec{v}_s + e^*\vec{A} = 0 , \qquad (2)$$

where $v_s$ is the Cooper-pair velocity. This also illustrates the fact that the force on the Cooper-pair inside the bulk is zero[2]. In a planar geometry, $v_s=2\omega$ and Equ. (2) gives immediately Equ. (1). As a consequence, even for a classical Type-I superconductor, the London field is believed to penetrate also the bulk of the superconductor[4]. This is in contrast to the Meissner-Ochselfeld effect, where a superconductor expels all magnetic fields from its interior. The London field was measured inside superconducting rings[3,5,6] as well as outside rotating superconducting discs[7-9] and agree well with Equ. (1). The London field measurement by Tate at el[5-6] was so precise as to allow the measurement of the Cooper-pair mass in Niobium. Contrary to theoretical predictions of $m^*/2m_e$ = 0.999992, she found a value of 1.000084(21), where $m_e$ is the electron mass. This anomaly is an active area of discussion which is still not resolved[4,10-12,36-37]. The authors recently conjectured that in addition to the classical London moment, also a gravitational analogue (a so-called gravitomagnetic London moment) would appear that explains Tate's measurement[13-14].

Using the quantum field theoretical approach to superconductivity, where photons gain mass due to the Higgs-mechanism via symmetry breaking, we will show that the London moment is directly related to the mass of the photon. We will then use the same formalism to express its gravitational analogue, the gravitomagnetic London moment, and to relate the gravitomagnetic field necessary to solve the Tate Cooper-pair mass anomaly to a massive graviton inside the superconductor.

**The London Moment Derived from Proca Equations**

Superconductivity is accompanied by breaking of gauge symmetry. In quantum field theory, this symmetry breaking leads to massive photons via the Higgs mechanism (for example see Ref. 15). In this case, the Maxwell equations transform to the so-called Proca equations (massive electromagnetism), which are given by

$$div\,\vec{E} = \frac{\rho}{\varepsilon_0} - \left(\frac{m_{Photon}c}{\hbar}\right)^2 \cdot \varphi$$

$$div\,\vec{B} = 0$$

$$rot\,\vec{E} = -\frac{\partial \vec{B}}{\partial t}$$

$$rot\,\vec{B} = \mu_0 \rho \vec{v} + \frac{1}{c^2}\frac{\partial \vec{E}}{\partial t} - \left(\frac{m_{Photon}c}{\hbar}\right)^2 \cdot \vec{A}$$

(3)

where $m_{Photon}$ is the Photon's mass. We can get both the Meissner-Ochsenfeld effect and the London moment by simply taking the curl of the 4$^{th}$ Proca equation. Neglecting the term coming from the displacement current, we get an equation for the magnetic field

$$\Delta \vec{B} - \frac{1}{\lambda_{Photon}^2}\vec{B} = \frac{2\vec{\omega}}{\lambda_L^2} \cdot \frac{m^*}{e^*},$$

(4)

where $\lambda_L$ is the London penetration depth. Following the motivation from Becker el al[1] and London[16], the London moment is developed by a net current that is lagging behind the positive lattice matrix; therefore, the Cooper-pair current density direction sign has to show in opposite direction than the angular velocity of the superconducting bulk. This is important as the London moment in all measurements, due to the negative charge of the Cooper-pair, shows in the same direction as the angular velocity. We can now solve Equ. (4) and express it for a one-dimensional case as

$$B = B_0 \cdot e^{-\frac{x}{\lambda_{Photon}}} - 2\omega \frac{m^*}{e^*}\left(\frac{\lambda_{Photon}}{\lambda_L}\right)^2.$$

(5)

Following quantum field theory, the photon wavelength is equal to the London penetration depth[15]. So we get the final result as

$$B = B_0 \cdot e^{-\frac{x}{\lambda_L}} - 2\frac{m^*}{e^*}\omega \ . \qquad (6)$$

The first part of Equ. (6) is of course the Meissner-Ochselfeld effect, $B_0$ is the externally applied magnetic field, and the second part is the London moment. In fact, the Proca equations with massive photons can be considered as a combination of both Maxwell and London equations[17].

**The Gravitomagnetic London Moment and the Graviton Mass**

In order to solve the Tate Cooper-pair mass anomaly, the authors proposed to add a gravitomagnetic field term to the London moment based on the conservation of the full canonical momentum[13],

$$\vec{B} = -\frac{2m^*}{e^*}\cdot\vec{\omega} - \frac{m^*}{e^*}\cdot\vec{B}_g \ . \qquad (7)$$

Gravitomagnetism is a weak field approximation to Einstein's general relativity theory, which is frequently used if space curvature effects can be neglected[18,19]. Classically, gravitomagnetic fields are very weak (such as the Earth's Lense-Thirring field currently measured by Gravity Probe-B[20]). However, the gravitomagnetic field required to solve the Tate anomaly is $\left|\vec{B}_g\right| = 2\vec{\omega}(\Delta m^*/m^*)$, where $\Delta m^*$ is the difference between the measured and the theoretically expected Cooper-pair mass in Tate's experiment[14]. Following our conjecture, $B_g$ is positive if the real Cooper-pair mass is the theoretical one ($m^*/2m_e = 0.999992$) and negative if the real Cooper-pair mass is the one measured by Tate ($m^*/2m_e = 1.000084$) where the negative $B_g$ would then correct the theoretical prediction.

Let us compare this value with the result that we expect classically from the gravitomagnetic field produced by a rotating ring[18], which can be calculated as

$$B_g = \frac{\mu_{0g}}{2r} \cdot \frac{dm}{dt} = \frac{4\pi G}{c^2} \cdot \frac{\rho \omega A}{2r^2}, \tag{8}$$

where $\rho$ is the density, $A$ the cross section area and $r$ the diameter of the rotating ring. Using Tate's geometry, we get a value that is 31 orders of magnitude below the value that we conjecture as a result from the rotation of the Niobium superconductor.

What could be the origin for such an effect? By doing a similar analysis of the Proca equations for gravitomagnetic fields in rotating superconductors as we did for electromagnetism, we see that the graviton mass could be ultimately responsible for such an amplification. This would not come as a surprise, as the photon mass in a superconductor due to gauge symmetry breaking is about $10^{-35}$ kg being some 34 orders of magnitude larger that the current upper limit of $10^{-69}$ kg in free space due to Heisenberg's uncertainty principle[21].

The Proca equations for gravitomagnetism can be derived straightforward from their electromagnetic counterparts as[22]

$$\begin{aligned}
\operatorname{div} \vec{g} &= -\frac{\rho_m}{\varepsilon_g} - \left(\frac{m_{Graviton} c}{\hbar}\right)^2 \cdot \varphi_g \\
\operatorname{div} \vec{B}_g &= 0 \\
\operatorname{rot} \vec{g} &= -\frac{\partial \vec{B}_g}{\partial t} \\
\operatorname{rot} \vec{B}_g &= -\mu_{0g} \rho_m \vec{v} + \frac{1}{c^2} \frac{\partial \vec{g}}{\partial t} - \left(\frac{m_{Graviton} c}{\hbar}\right)^2 \cdot \vec{A}_g
\end{aligned} \tag{9}$$

Similarly, we can now take the curl of the 4$^{th}$ equation to analyse the gravitomagnetic Meissner-Ochsenfeld and London moment. In this case we don not have to reverse the current sign as there are no counter currents between positive ions (from the lattice) and negative Cooper-pairs flowing. In the gravitomagnetic case, Becker's argument[1] clearly does not hold and we expect a gravitomagnetic field being generated not only in a thin penetration depth close to the surface but everywhere inside the superconducting bulk.

For our analysis, we also have to use the gravitomagnetic analogue of the London penetration depth[23]. This can be derived simply from the 2nd London equation for gravitomagnetic fields $rot\, j_s = -n_s e B_g$ together with $rot\, \vec{B}_g = -\mu_{0g} j_s$, which yields the gravitomagnetic London penetration depth as

$$\lambda_{Lg} = i\sqrt{\frac{1}{\mu_{0g} n_s m^*}} \,. \tag{10}$$

Note that this penetration depth is a complex number contrary to the real number in the electromagnetic case. That goes also along our argument above that we expect the generation of gravitomagnetic fields not inside a thin layer; being a complex penetration depth, gravitomagnetic fields would penetrate the superconducting bulk with an oscillating signature.

Proceeding with the Proca equation as we did in the electromagnetic case, we arrive at

$$B_g = B_{g0} \cdot e^{-\frac{x}{\lambda_g}} - 2\omega \left(\frac{\lambda_g}{\lambda_{Lg}}\right)^2 , \tag{11}$$

where we neglected again the displacement term and used the graviton wavelength $\lambda_g = \hbar/m_g.c$ with $m_g$ as the graviton mass. Here we did not initially equal the graviton wavelength and the gravitomagnetic London penetration depth as the Higgs mechanism for gravitons is certainly different from the one with photons. We immediately recognize that the gravitomagnetic London moment due to the graviton mass in Equ. (11) is directly proportional to $-2\omega$, just as we predicted it to solve Tate's Cooper-pair mass anomaly[14] with $B_g = -2\omega(\Delta m^*/m^*)$. As $B_g$ is negative, it appears that Tate measured indeed the right Cooper-pair mass and that the theoretical prediction has to be modified due to the additional large gravitomagnetic term.

If our conjecture is correct, we can now use Tate's measurement to actually express the mass of the graviton in a Niobium superconductor as

$$m_g = i \cdot \sqrt{\frac{\mu_{0g} n_s m^{*2} \hbar^2}{c^2 \Delta m^*}} = i \cdot 4.61 \times 10^{-55} \; kg \;\;. \tag{12}$$

Note that the graviton mass predicted is a complex number. This is in agreement with a recent theoretical assessment by Novello et al[24] who showed that the graviton mass must be complex in a de-Sitter background (positive cosmological constant) which is currently observed in the universe. The corresponding graviton wavelength is then

$$\lambda_g = \frac{\hbar}{m_g c} = i \cdot 7.63 \times 10^{11} \; m \;\;.^1 \tag{13}$$

How does our graviton mass in Equ. (12) compare with experimental upper-limits for gravitons in free space? As the cosmological constant, $\Lambda$, in Einstein's field equations is related to the wavelength of a graviton ($\lambda_g^{-2} \cong -\Lambda$, see e.g. Ref. 24), we can use the presently established measurements for the cosmological constant[25] in the universe $\Lambda_0 = (1.29 \pm 0.23) \times 10^{-52}$ m$^{-2}$. Therefore, the upper limit of the complex graviton mass in free space is $m_{g0}$~i.10$^{-69}$ kg. This is the same upper limit (in real numbers) as we got for the photon mass in free space. If both masses are related in free space, why should then the graviton mass not be able to change if the photon acquires mass in a superconductor?

---

[1] For comparison, the real part of this wavelength is very similar to the distance Sun-Jupiter in the solar system.

**Discussion**

We saw that if the graviton mass inside a superconductor is $10^{-55}$ kg, it can explain the Cooper-pair mass anomaly from Tate. This is "only" 14 orders of magnitude above its presently accepted upper limit in free-space. In a recent assessment, Modanese[27] calculated the cosmological constant inside a superconductor taking into account the contribution of the Ginzburg-Landau wave function $\psi_{GL}$ to the Lagrangian. He found that in the case of a Pb superconductor, the cosmological constant should be on the order of $10^{-39}$ m$^{-2}$. That would lead to a complex graviton mass of i.$10^{-62}$ kg, coming closer to our estimate of i.$10^{-55}$ kg in a Nb superconductor.

What would happen if the graviton wavelength would be equal to the gravitomagnetic London penetration depth (as we have it in the case of electromagnetic fields)? Then the gravitomagnetic London moment would be

$$\vec{B}_g = -2\vec{\omega} . \qquad (25)$$

Looking at the conservation of the canonical moment as expressed by Equ. (7), this case also corresponds to a vanishing classical London moment in the superconductor – and correspondingly to an infinite photon mass. Is that possible?

Verheijen et al performed[7,8] London moment measurements of Pb, YBa$_2$Cu$_3$O$_{7-x}$ (YBCO) and BaPb$_{0.8}$Bi$_{0.2}$O$_3$ (BPBO) superconductors. Surprisingly, they found that the slope (magnetic field versus angular velocity) for the BPBO is 25% less that its predicted value (-2m$^*$/e$^*$ ω) while the YBCO and Pb superconductors agreed to within 10% of their expected values. They explained it by the partial magnetic shielding property (only 91% at 4.2 K) common to ceramic high-T$_c$ superconductors that is may be related to inhomogeneous composition and weak coupling between grains.

Looking at our Proca equation assessment, the slope of the London moment can only change if the photon wavelength is different from the London penetration depth (see Equ. (5)). It is maybe possible to increase the photon mass and correspondingly decrease the graviton mass by properly designing the composition of the superconductor. Obviously, that needs further theoretical and most important experimental verification. However, recent claims on gravity waves produced by HV discharges on ceramic superconducting electrodes were also reported to crucially depend on the sintered microstructure of the superconductor[28]. Equ. (25) in fact is only possible assuming an infinite photon mass and a graviton mass of $10^{-57}$ kg, which is still 12 orders of magnitude above its present upper limit in free space.

**Conclusion**

Using the Proca equations, we could express the classical London moment as a direct consequence of the photon mass inside a superconductor. Doing a similar assessment for gravitational fields, we could express a gravitomagnetic London moment that depends on the angular velocity, gravitomagnetic penetration depth and the graviton wavelength. Following a recent suggestion by the authors to solve the Tate Cooper-pair mass anomaly by considering a non-classical gravitomagnetic field inside the superconductor, we can calculate the mass of the graviton needed to solve the anomaly inside a Nb superconductor. The value of $10^{-55}$ kg is about 14 orders of magnitude higher than its presently accepted upper limit in free space. However, also the photon mass is greatly increased inside a superconductor due to the Higgs mechanism and symmetry breaking. An equivalent mechanism could take part for the graviton as well. This mechanism most certainly depends on the answer to another question: Since Cooper pairs have anomalous mass excess, and all rest mass comes from the Higgs mechanism, and since rest mass interacts through gravity, what is the relationship between gravity and the Higgs mechanism in a superconductor? Some attempts to answer parts of that question can be found in the literature[29-35], but the final quantum theory of gravity is still not completed. London moment measurements on high-$T_c$ superconductors as well as claims on the generation of gravitational waves suggest that the composition of the superconductor could play a crucial role in the ratio between the photon and graviton masses with respect to their penetration depths. That could in principle lead to large gravitational effects that should be detectable in a laboratory environment, and settle the experimental ground for the understanding of gravity at quantum level.

**References**

[1] Becker, R., Heller, G., and Sauter, F., "Über die Stromverteilung in einer Supraleitenden Kugel", *Z. Physik*, **85**, 1933, pp. 772-787

[2] Rystephanick, R.G., "On the London Moment in Rotating Superconducting Cylinders", *Can. J. Phys.*, **51**, 1973, 789-794

[3] Hildebrandt, A.F., "Magnetic Field of a Rotating Superconductor", *Physical Review Letters*, **12**(8), 1964, pp. 190-191

[4] Capellmann, H., "Rotating Superconductors: Ginzburg-Landau Equations", *European Physical Journal B*, **25**, 2002, pp. 25-30

[5] Tate, J., Cabrera, B., Felch, S.B., Anderson, J.T., "Precise Determination of the Cooper-Pair Mass", *Physical Review Letters*, **62**(8), 1989, pp 845-848

[6] Tate, J., Cabrera, B., Felch, S.B., Anderson, J.T., "Determination of the Cooper-Pair Mass in Niobium", *Physical Review B*, **42**(13), 1990, pp 7885-7893

[7] Verheijen, A.A., van Ruitenbeek, J.M., de Bruyn Ouboter, R., and de Jongh, L.J., "The London Moment for High Temperature Superconductors", *Physica B*, **165-166**, 1990, pp. 1181-1182

[8] Verheijen, A.A., van Ruitenbeek, J.M., de Bruyn Ouboter, R., and de Jongh, L.J., "Measurement of the London Moment for Two High Temperature Superconductors", *Nature*, **345**, 1990, pp. 418-419

[9] Sanzari, M.A., Cui, H.L., and Karwacki, F., "London Moment for Heavy-Fermion Superconductors", *Appl. Phys. Lett.*, **68**(26), 1996, pp. 3802-3804

[10] Liu, M., "Rotating Superconductors and the Frame-Independent London Equation", *Physical Review Letters*, **81**(15), 1998, pp. 3223-3226

[11]Jiang, Y., and Liu, M. "Rotating Superconductors and the London Moment: Thermodynamics versus Microscopics", *Physical Review B*, **63**, 2001, 184506

[12]Berger, J., "Nonlinearity of the Field Induced by a Rotating Superconducting Shell", 2004

[13]Tajmar, M., and de Matos, C.J., "Gravitomagnetic Field of a Rotating Superconductor and of a Rotating Superfluid", *Physica C*, **385**(4), 2003, pp. 551-554

[14]Tajmar, M., and de Matos, C.J., "Extended Analysis of Gravitomagnetic Fields in Rotating Superconductors and Superfluids", Physica C, submitted (2004)

[15]Ryder, L.H., "Quantum Field Theory", Cambridge University Press, 2$^{nd}$ Edition, 1996, pp. 296-298

[16]London, F., "Superfluids", John Wiley & Sons, New York, 1950

[17]Sternberg, S., "On the London Equations", Proc. Natl. Acad. Sci., **89**, 1992, pp. 10673-10675

[18]Forward, R.L., "General Relativity for the Experimentalist", *Proceedings of the IRE*, 1961, pp. 892-586

[19]Tajmar, M., de Matos, C.J., "Coupling of Electromagnetism and Gravitation in the Weak Field Approximation", *Journal of Theoretics*, **3**(1), February 2001 (also gr-qc/0003011)

[20]Buchman, S., et al, "The Gravity Probe B Relativity Mission", *Adv. Space Res.*, **25**(6), 2000, pp.1177

[21]Tu, L., Luo, J., and Gillies, G.T., "The Mass of the Photon", *Rep. Prog. Phys.*, **68**, 2005, pp. 77–130

[22]Argyris, J., Ciubotariu, C., "Massive Graviton in General Relativity", Austr. J. Phys., **50**, 1997, pp. 879-91

[23]de Matos, C.J., and Tajmar, M., "Gravito-Electromagnetic Properties of Superconductors - A Brief Review", *Los Alamos Physics Archive*, cond-mat/0406761, 2004

[24]Novello, M., and Neves, R.P., "The Mass of the Graviton and the Cosmological Constant", *Classical and Quantum Gravity*, **20**, 2003, L67-L73

[25]Spergel, D.N., et al., *Astrophy. J. Suppl.*, **148**, 2003, pp. 175 and references therein

[26]Pascual-Sanchez, J.F., "On the (Non) Existence of Several Gravitomagnetic Effets", gr-qc/9906086, 1999

[27]Modanese, G., "Local Contribution of a Quantum Condensate to the Vacuum Energy Density", *Mod. Phys. Lett. A*, **18**(10), 2003, pp. 683-690

[28]Podkletnov, E., and Modanese, G., "Investigations of HV Discharges in Low Pressure Gases through Large Ceramic Superconducting Electrodes," *Journal of Low Temperature Physics*, **132**(3/4), 2003, pp. 239-259

[29]Sardanashvily, G. A., "Gauge Gravitation Theory. What is the Geometry of the World?", *Los Alamos Physics Archive*, gr-qc/9410045, 1994

[30]Sardanashvily, G. A., "Gravity as a Higgs Field. I. Geometric Equivalence Principle", *Los Alamos Physics Archive*, gr-qc/9405013, 1994

[31]Sardanashvily, G. A., "Gravity as a Higgs Field. II. Fermion-Gravitation Complex", *Los Alamos Physics Archive*, gr-qc/9407032, 1994

[32]Sardanashvily, G. A., "Gravity as a Higgs Field. III. Nongravitational Deviations of Gravitational Fields", *Los Alamos Physics Archive*, gr-qc/941103, 1994

[33]Bluhm, R., Kostelecky, V. A., "Spontaneous Lorentz violation, Nambu-Goldstone Modes, and Gravity", *Los Alamos Physics Archive*, hep-th/0412320, 2004

[34]Smith, F. D. T., "SU(3) X SU(2) X U(1), higgs, and Gravity from *Spin*(0,8) Clifford Algebra *Cl*(0,8)", *Los Alamos Physics Archive*, hep-th/9402003, 1994

[35]Smith, F. D. T., "Higgs and Fermions in $D_4 - D_5 - E_6$ Model based on *Cl*(0,8) Clifford Algebra", *Los Alamos Physics Archive*, hep-th/9403007, 1994

[36]Capelle, K., Gross, E. K. U., "Relativistic Framework for the Microscopic Theories of Superconductivity. I. The Dirac Equation for Superconductors", *Physical Review B*, **59**, 10, 1999

[37]Capelle, K., Gross, E. K. U., "Relativistic Framework for the Microscopic Theories of Superconductivity. II. The Pauli Equation for Superconductors", *Physical Review B*, **59**, 10, 1999